\documentclass[manuscript]{acmart}

\usepackage{amsmath}
\usepackage{booktabs}
\usepackage{graphicx}

\setcopyright{none}
\renewcommand\footnotetextcopyrightpermission[1]{}
\acmDOI{}
\acmISBN{}

\begin{document}

\title{\textit{NeuroDivSim}: An Interactive Tool for Model-Based Reflection on Cognitive Diversity in Interface Design}

\author{Eske Beckefeld}
\affiliation{%
  \institution{University of Applied Sciences Bremen}
  \city{Bremen}
  \country{Germany}}
\email{eske.beckefeld@hs-bremen.de}

\author{Henrik H.J. Detjen}
\affiliation{%
  \institution{Fraunhofer Institute for Digital Medicine MEVIS}
  \city{Bremen}
  \country{Germany}}
\affiliation{%
  \institution{University of Applied Sciences Bremen}
  \city{Bremen}
  \country{Germany}}
\email{henrik.Detjen@mevis.fraunhofer.de}

\begin{abstract}
Recent approaches to simulated and synthetic users offer new ways to support design, but raise questions about how computational representations of users should contribute to design practice. We present \textit{NeuroDivSim}, an interactive tool that explores simulation as an inspectable mechanism for reflecting on cognitive diversity during design and prototyping. Rather than using an LLM to act as a simulated user, NeuroDivSim uses generative AI to construct inspectable task, interface, and environment models from a usage scenario. After human review, these models are combined with explicit cognitive reference configurations and processed through deterministic simulation. This enables designers to hold a modeled usage situation constant while varying cognitive assumptions and tracing their consequences to interaction steps and rule-based design recommendations. We further report an exploratory pilot evaluation (\(N=10\)) that provided formative insights into how participants engaged with the workflow and informed subsequent refinements to the presentation of models, simulation results, and recommendations.
\end{abstract}
\begin{teaserfigure}
  \includegraphics[width=\textwidth]{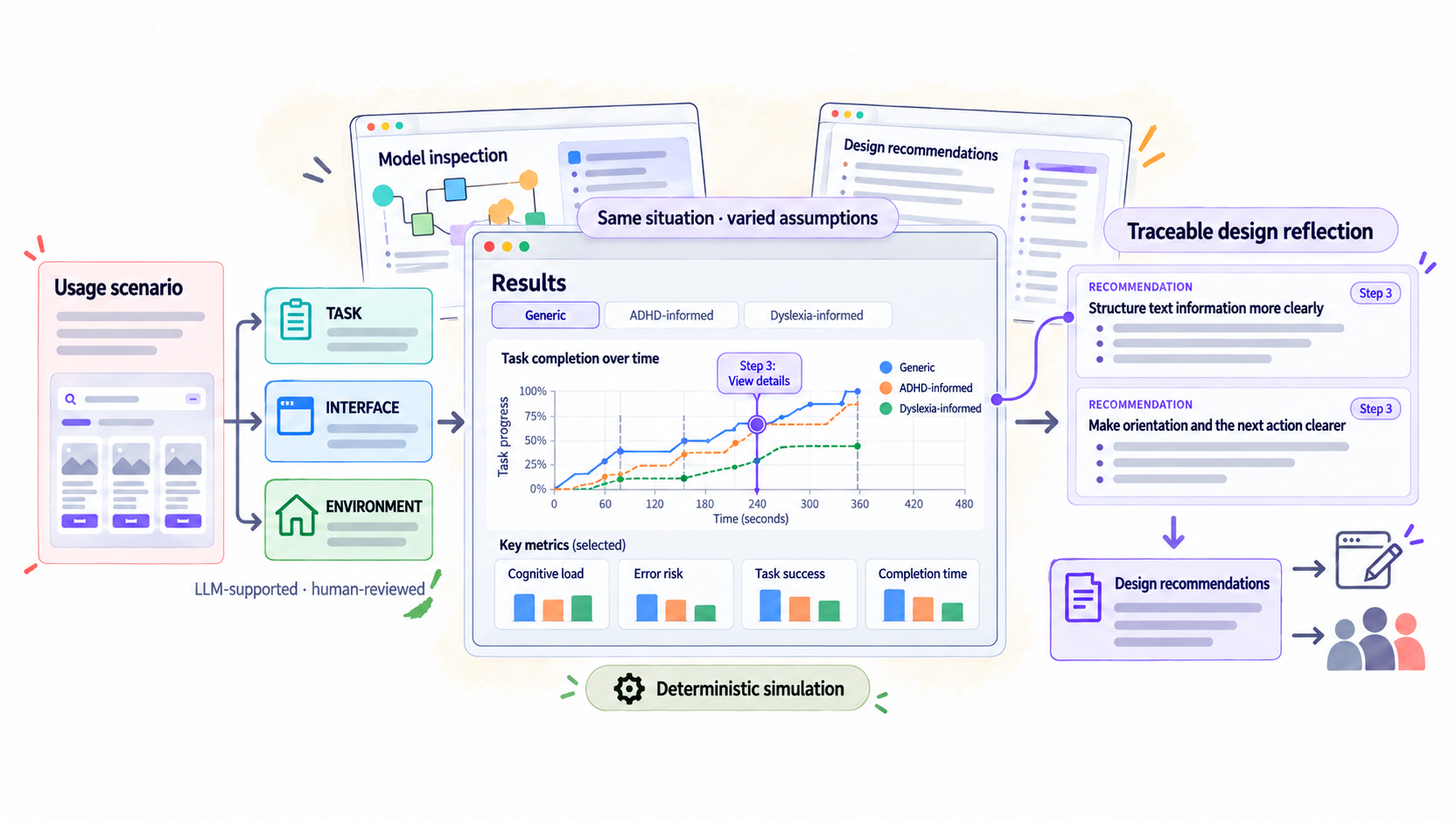}
  \vspace{-1.5cm}
  \caption{Our proposed workflow and the prototypical open source implementation \textit{NeuroDivSim} (\url{https://github.com/eskebfd/neurodivsim}) helps designers to reflect on cognitive diversity during early design phases.}
  \Description{Our proposed workflow and the prototypical open source implementation \textit{NeuroDivSim} helps designers to reflect on cognitive diversity during early design phases.}
  \label{fig:teaser}
\end{teaserfigure}

\maketitle

\section{Introduction}
\label{sec:introduction}

Recent advances in generative AI have expanded computational approaches to representing and simulating users toward generative agents, AI-generated personas, and simulated users that can produce user-like behavior or feedback \cite{10.1145/3586183.3606763,10.1145/3772318.3790608,10.1145/3613904.3642481}. While such approaches can provide additional perspectives during design, using simulated users as representations of human participants raises concerns about validation, stereotyping, and representational bias \cite{10.1145/3772318.3790608,10.1145/3745900.3746108}. These concerns are particularly relevant for heterogeneous populations that cannot be adequately represented by generalized user models.

Neurodiversity provides a salient case for this tension. HCI research emphasizes substantial variation within neurodivergent populations and the importance of participatory and strengths based approaches for understanding users' needs and experiences \cite{10.1145/3711106,10.1145/3613904.3642424,10.1145/3772318.3790697}. Yet designers must still translate knowledge from research, guidelines, and user involvement into concrete decisions about interfaces, tasks, and contexts between opportunities for empirical evaluation. This creates a need for approaches that support reasoning about cognitive diversity while designs remain open to revision.

Computational user models can support such reasoning by formalizing assumptions and enabling systematic comparison of their modeled consequences \cite{10.1145/3491102.3517739,10.1145/3613904.3642063}. Recent systems extend this idea through LLM based user simulation. SimUser, for example, uses an LLM based user agent to simulate interaction and generate usability feedback for specified users and contexts \cite{10.1145/3613904.3642481}. We investigate a complementary question: can simulation support designers without requiring a generative agent to act as a surrogate for a human user?

We present \textit{NeuroDivSim}, an interactive tool that uses simulation as an inspectable mechanism for reflection on cognitive diversity during design and prototyping (cf. \autoref{fig:teaser}). An LLM translates a natural language usage scenario and optional interface screenshot into inspectable task, interface, and environment models, which are combined with predefined cognitive reference configurations. Subsequent processing is deterministic, allowing designers to hold the modeled usage situation constant while varying cognitive assumptions and tracing their consequences to interaction steps and rule based design recommendations.

This architecture supports counterfactual design reasoning by making the consequences of selected cognitive assumptions explicit and available for scrutiny while a design can still be revised. Rather than predicting how neurodivergent people will behave, NeuroDivSim uses simulation as a tool for reflection on modeled assumptions.

This paper makes three contributions. First, we frame simulation as an inspectable reflection mechanism for reasoning about cognitive diversity rather than as a surrogate for empirical participants. Second, NeuroDivSim instantiates this framing through LLM supported scenario modeling, human inspection, deterministic simulation, and traceable design recommendations. Third, an exploratory pilot study with ten prospective users provides formative evidence about engagement with the workflow and identifies requirements for presenting its models, simulation results, and recommendations.

\section{Related Work}
\label{sec:related-work}

\subsection{Neurodiversity and Ability-Based Design}
\label{sec:neurodiverse-design}

Neurodiversity describes variation in neurocognitive functioning and challenges a single normative way of thinking, learning, or interacting \cite{chapman2021neurodiversity,10.1145/3711106}. Characteristics related to attention, working memory, information processing, and reading can combine with interface and task demands. Prior work reports attention-regulation, task-management, and distraction-related challenges for people with ADHD \cite{brown2018,10.1145/3772318.3790697,10.1145/3772363.3799383}, and interactions between reading characteristics and text-heavy interfaces for people with dyslexia \cite{snowling2022,10.1145/3627508.3638292,10.1145/3800645.3813089}.

Such findings do not imply uniform requirements within diagnoses: neurodivergent people differ in characteristics, strategies, contexts, and access needs \cite{10.1145/3711106}. Critical reviews show how top-down, deficit-oriented HCI can marginalize neurodivergent interests \cite{spiel2021purpose}, while accessibility scholarship calls for disability studies and disabled participants to shape both problems and methods \cite{mankoff2010disability,mack2021accessibility}. Participatory and strengths-based work therefore avoids generalized deficits \cite{10.1145/3613904.3642424,10.1145/3772318.3790697}. Ability-Based Design likewise foregrounds what people can do in context and places responsibility on systems to accommodate heterogeneous abilities \cite{wobbrock2011ability}. This motivates representing interaction-relevant dimensions independently rather than treating diagnoses as behavioral specifications. Models cannot establish what an interaction means to a person; NeuroDivSim instead supports provisional decisions between empirical engagements by turning prior evidence and stakeholder knowledge into explicit, revisable questions.

\subsection{Model-Based Inspection and Reflection}
\label{sec:simulation-reflection}

HCI has long evaluated designs without treating an analytical model as a participant. Cognitive walkthroughs inspect task steps using theory-guided questions \cite{wharton1992walkthrough}, while GOMS-based methods formalize tasks and operators to predict aspects of interaction performance \cite{john1996goms,john2004predictive}. Recent computational models incorporate cognitive constraints, adaptation, and environmental conditions \cite{10.1145/3491102.3517739,10.1145/3613904.3642063}, including attention allocation in dynamic tasks \cite{gasse2025attention,lorenz2026creyes}. Their growing relevance to work contexts has also motivated a dedicated research agenda \cite{10.1145/3805029.3818274}. Such models enable counterfactual reasoning by holding parts of an interaction constant while varying selected assumptions \cite{10.1145/3491102.3517739}.

NeuroDivSim combines inspection and counterfactual modeling. Like a walkthrough, it structures analysis around task steps; unlike a conventional walkthrough, it compares parameterized cognitive and contextual assumptions across the same steps. Its present purpose is not calibrated prediction, but identifying where assumptions produce a modeled concern. This connects to reflective design \cite{sengers2005reflective} and design rationale \cite{10.1145/1182475.1182511}. Designers can question attributes, parameters, thresholds, and rules and compare runs; prior work nevertheless cautions that explanations may aid intelligibility without ensuring correct interpretation \cite{lim2009why,kaur2020interpreting}. Because AI's workflow role affects engagement and agency \cite{10.1145/3706598.3713649}, NeuroDivSim uses an LLM only to propose an editable scenario representation; subsequent comparisons are deterministic.

\subsection{Synthetic Users and the Boundaries of Simulation}
\label{sec:simulation-boundaries}

Generative AI has expanded user representations toward agents, AI-generated personas, and synthetic users \cite{10.1145/3586183.3606763,10.1145/3772318.3790608}. SimUser generates usability feedback through an LLM-based user agent \cite{10.1145/3613904.3642481}, while DiverseClaire uses LLM-based learner personas to assess educational materials \cite{10.1145/3770761.3777340}. Such approaches can add perspectives, but raise concerns about validity, stereotyping, representational bias, transparency, and substitution for empirical research \cite{10.1145/3772318.3790608,10.1145/3745900.3746108}.

Related critiques concern disability simulations and empathy exercises. Temporary approximations may be mistaken for lived experience, privilege designers' interpretations, and even reduce judged capabilities of disabled people \cite{bennett2019empathy,silverman2015stumbling}. VIP-Sim addresses this tension through participatory development \cite{raedler2025vipsim}; empirical work also shows that guideline judgments need not match problems experienced by disabled users \cite{petrie2007relationship}. NeuroDivSim therefore neither approximates experience nor speaks as a neurodivergent person. Its LLM constructs an editable usage model rather than testimony; explicit configurations and rules expose how assumptions propagate through a fixed task. Simulation supports model inspection, not surrogate participation or claims about individual behavior.

\section{Design of \textit{NeuroDivSim}}
\label{sec:design}

\textit{NeuroDivSim} operationalizes cognitive diversity as an explicit and inspectable modeling problem. The system separates scenario interpretation from simulation: an LLM first transforms an informal usage description into structured representations of the task, interface, and environment. Designers can inspect and revise these representations before they enter the simulation. Cognitive assumptions are represented independently through reference configurations and are combined with the reviewed scenario models only during deterministic processing. This separation allows assumptions introduced by generative interpretation, cognitive modeling, and simulation logic to remain distinguishable throughout the workflow.

\begin{figure}[H]
    \centering
    \makebox[\textwidth][c]{%
        \includegraphics[width=1.15\textwidth]{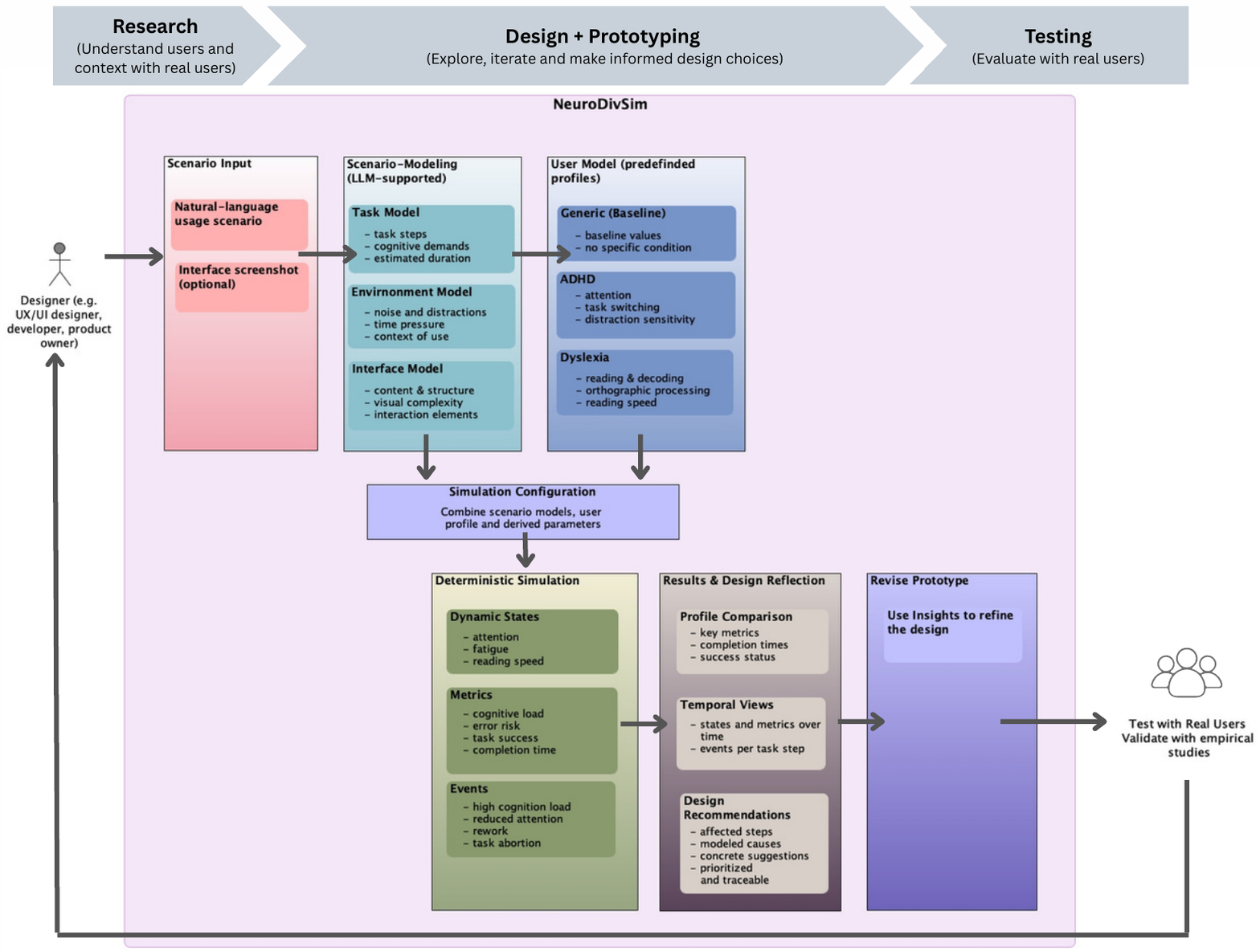}
    }
    \caption{NeuroDivSim workflow from scenario modeling to design reflection.}
    \Description{The figure places NeuroDivSim within an iterative process from research to design and prototyping to testing. A natural language scenario and optional interface screenshot are transformed into task, environment, and interface models. These models are combined with predefined cognitive reference configurations and processed by a deterministic simulation. Simulation states, metrics, and events are organized into comparisons, temporal views, and design recommendations that support prototype revision before testing with real users.}
    \label{fig:workflow}
\end{figure}

Figure~\ref{fig:workflow} summarizes the resulting process and its position between research, design and prototyping, and subsequent empirical testing.

A detailed interaction workflow, covering the complete process from scenario specification and model construction to simulation and result exploration, is provided in Appendix~\ref{app:interaction-workflow}. Additional interface screenshots in Appendix~\ref{app:interface-examples} illustrate selected stages of this workflow, including scenario specification, interaction flow and requirement inspection, simulation results, and design recommendations.

%To support inspection of the implemented workflow, an anonymized version of the NeuroDivSim prototype and source code is available at \url{https://osf.io/f4z8e/overview?view_only=6fa6e06bfba146639e94cd6a6006f37d}.
To support inspection of the implemented workflow, we provide the \textit{NeuroDivSim} source code at \url{https://github.com/eskebfd/neurodivsim}.

\paragraph{Running example.}
To illustrate how these components interact, we use a hotel-booking scenario throughout this section. A traveler needs to book suitable accommodation for a three-day business trip to Berlin. The booking should fit the available budget, meeting schedule, and cancellation conditions. The modeled interface contains a search form, a result list with hotel cards, filters, and detail pages containing information about room options, additional costs, and cancellation terms. The environmental description specifies that the task is performed on a laptop in an open-plan office, with background conversations, occasional notifications, and moderate time pressure because the booking should be completed before the next meeting. This example is used to illustrate the transformation from an informal usage situation to explicit models, simulation outcomes, and design recommendations. Corresponding interface views are provided in Appendix~\ref{app:interface-examples}.

\subsection{Formalizing the Usage Situation}
\label{sec:model-construction}

NeuroDivSim represents a usage situation through separate task, interface, and environment models. This separation reflects the assumption that interaction demands do not arise from user characteristics alone, but from their relation to a particular task, interface, and context.

The task model combines global task characteristics with a stepwise decomposition of the interaction. Global attributes include task complexity, number of steps, reading demand, input demand, and memory demand. The decomposition is conceptually informed by Hierarchical Task Analysis and GOMS \cite{annett2004hta,john1996goms}, which provide established approaches for structuring tasks and user actions. NeuroDivSim does not implement either method in its full formal sense; instead, it adopts their general principle of decomposing an interaction into explicit task steps. These steps provide the structure to which subsequent simulation events and recommendations can be associated.

The interface model describes properties that may contribute to cognitive interaction demands, including text volume, sentence length, word difficulty, technical terminology, visual clutter, navigation complexity, accessibility support, and feedback quality. The environment model independently represents contextual influences including noise, distractions, time pressure, and context stability. Keeping these dimensions separate makes it possible to distinguish, for example, a reading intensive task from a textually complex interface or an otherwise manageable interaction performed under distracting conditions.

Model attributes are represented on a common 0--100 scale. The LLM maps information from the natural language scenario and optional screenshot to these structured attributes and can provide rationales linking generated values to evidence in the input. Importantly, this stage produces simulation inputs rather than simulation outcomes. Designers can inspect and revise generated attributes and task steps before confirming the models.

In the running example, NeuroDivSim decomposes the hotel-booking task into explicit interaction steps. These include entering the search criteria, filtering the results, comparing offers, viewing hotel details, and subsequent booking steps. Each step is associated with actions, an estimated duration, and cognitive requirements. Entering the search criteria, for example, involves perceiving the form, typing the required information, and submitting it, whereas comparing offers requires information to be perceived and compared and is associated with decision-making and sustained focus.

The same scenario is translated into inspectable model attributes. The task model, for example, represents properties such as task complexity, reading demand, input demand, and memory demand on the shared 0--100 scale. Interface and environmental properties remain separate. This distinction is relevant in the hotel-booking example because demands during the comparison of hotel offers may originate from the task of comparing several alternatives, from the amount and structure of information presented by the interface, or from contextual factors such as background conversations and time pressure. NeuroDivSim therefore retains these as separate sources of modeled demand rather than collapsing them into a single difficulty score.

After model confirmation, the LLM is removed from the processing pipeline.

\subsection{Representing Cognitive Assumptions}
\label{sec:reference-profiles}

Cognitive assumptions are represented separately from the modeled usage situation. NeuroDivSim currently provides a generic comparison configuration together with ADHD informed and dyslexia informed reference configurations. These configurations are not intended as representations of individual people or population averages. Instead, they provide predefined combinations of cognitive parameters that can be applied to the same modeled situation.

The shared parameter space includes reading difficulty, attention stability, distraction sensitivity, task switching difficulty, and working memory stability. Further simulation parameters operationalize constructs including decoding, inhibition, vigilance, and reaction variability. Using a shared parameter space is methodologically important because differences between configurations result from changes to explicit parameter values rather than from different underlying simulation structures.

The selection and directional interpretation of these dimensions were informed by literature on ADHD and dyslexia \cite{brown2018,10.1145/3772318.3790697,10.1145/3772363.3799383,snowling2022,10.1145/3627508.3638292,10.1145/3800645.3813089}. Their numerical values, however, are heuristic operationalizations and should not be interpreted as empirically estimated characteristics of diagnostic populations. This distinction allows literature informed assumptions to be made computationally explicit without claiming behavioral prediction.

In the running example, the confirmed hotel-booking task, interface, and environment models remain identical when different reference configurations are selected. Selecting the ADHD informed configuration therefore does not change the booking scenario or its task decomposition. Instead, it changes only the cognitive parameters applied to that same modeled situation. This allows differences in subsequent simulation outcomes to be attributed to the explicitly varied assumptions while holding the remaining scenario representation constant.

\subsection{Deterministic Cognitive Simulation}
\label{sec:deterministic-simulation}

The reviewed scenario models and a selected cognitive reference configuration form the complete input to the simulation. NeuroDivSim first derives intermediate parameters from related attributes. These represent interaction demands such as text complexity, navigation effort, decoding effort, memory demand, and sustained attention load. The separation between source attributes and derived parameters preserves the origin of the factors entering the simulation and allows individual relationships to be inspected or replaced.

Simulation proceeds sequentially through the modeled task steps at one second intervals. Rather than assigning a single score to the complete scenario, NeuroDivSim maintains dynamic cognitive states across the interaction. These include attention, fatigue, and reading speed. Their development depends on the current task step and its interaction with interface, environment, and cognitive parameters. Consequently, modeled demands can accumulate or change over the course of an interaction rather than remaining constant across all steps.

From these states and parameters, the simulation derives cognitive load, error risk, task success, completion efficiency, completion time, and time limit risk. Threshold based conditions additionally produce events such as elevated cognitive load, reduced attention, or rework. Events retain their temporal and task context, allowing an aggregate difference between reference configurations to be traced back to the interaction steps and modeled conditions from which it emerged.

Applied to the running example, the same confirmed hotel-booking models can be simulated with each cognitive reference configuration while all task, interface, and environmental assumptions remain unchanged. In the illustrated simulation run, the ADHD informed configuration results in a simulated completion time of 399 seconds, compared with 389 seconds for the dyslexia informed configuration and 373 seconds for the generic comparison configuration. The corresponding aggregate error-risk values are 88.75, 86.72, and 81.07. These values are model outputs rather than measured user data and should not be interpreted as predictions of how people with ADHD or dyslexia would perform the booking task. Instead, they expose how changes to the predefined cognitive parameters propagate through the same computational representation of the interaction.

More importantly, NeuroDivSim retains the stepwise development underlying these aggregate results. Figure~\ref{fig:example-simulation-results} in Appendix~\ref{app:interface-examples} illustrates this for the ADHD informed reference configuration. The results view plots cognitive load, error risk, task success score, and completion efficiency across the individual interaction steps. A second temporal view shows the corresponding development of dynamic simulation states, including attention, fatigue, reading speed, and progress rate. Threshold based events are displayed as markers at the interaction steps at which the corresponding conditions are reached.

In the illustrated run, these trajectories show that the aggregate result does not emerge uniformly across the task. Attention decreases over the interaction while fatigue increases, and changes in error risk and completion efficiency can be inspected in relation to specific booking steps. Event markers additionally identify where conditions such as elevated error risk, strongly reduced attention, critical time pressure, or required rework occur. The visualization therefore allows a designer to move from an aggregate difference between configurations back to the individual task steps and modeled conditions that contributed to it.

All computations after model confirmation are deterministic. Given identical task, interface, environment, and cognitive inputs, NeuroDivSim produces identical state trajectories, metrics, and events. A comparison between configurations therefore changes only the explicitly varied cognitive assumptions. The simulation does not claim that the resulting values predict the behavior of an individual user; they represent consequences of the assumptions encoded in the model.

\subsection{From Simulation States to Design Recommendations}
\label{sec:design-reflection}

NeuroDivSim connects simulation outcomes to design reflection through a rule based recommendation layer. Rules evaluate combinations of task properties, interface attributes, cognitive parameters, simulation states, metrics, and events. When their conditions are met, they identify a potentially demanding interaction step and associate it with a corresponding design response.

Recommendations therefore remain linked to the conditions that triggered them. Each recommendation retains the affected task step, relevant modeled factors, an explanation of the detected relationship, and a concrete design suggestion. The LLM does not generate these recommendations. This distinction prevents variation in generative output from being introduced after designers have reviewed the scenario representation.

The running example illustrates this traceability at the level of an individual interaction step. For the ADHD informed configuration, Step~1 requires the traveler to fill in the search form with the required details. When the simulation indicates a notable decrease in attention at this step, the recommendation layer can trigger the recommendation ``Make orientation and the next action clearer.'' The recommendation remains associated with Step~1 and exposes the observation and rationale that led to it. Concrete suggested changes include visually emphasizing the primary next action, reducing competing elements, making the current processing status visible, and grouping related options more clearly.

The same interaction step can produce a different recommendation when another modeled condition is involved. Elevated error risk during Step~1 can trigger ``Prevent errors at this point.'' In the illustrated example, the associated design responses include clearly marking required fields, validating inputs directly, formulating error messages concretely and actionably, and summarizing critical inputs before submission. A reading-related demand at the same step can instead trigger ``Structure text information more clearly,'' with suggestions such as shortening descriptive text, introducing subheadings or visual grouping, and making detailed information optionally expandable.

The recommendation mechanism therefore does not associate design advice with a diagnostic label in isolation. In the example, the recommendation is the end of a traceable chain connecting a specific booking step, properties of the modeled task and interface, the selected cognitive assumptions, and the resulting simulation states or events. This allows a designer to inspect not only what change is suggested, but also why the modeled interaction produced that suggestion. The corresponding recommendation view is shown in Figure~\ref{fig:example-recommendations} in Appendix~\ref{app:interface-examples}.

The resulting chain from scenario evidence to explicit model attributes, cognitive assumptions, simulation states and events, and finally design recommendations provides the basis for inspectability in NeuroDivSim. Rather than presenting a synthetic user's judgment, the system exposes why a particular interaction step was flagged under a particular set of assumptions. Designers can modify the interface, scenario model, or selected assumptions and repeat the simulation to examine how these changes propagate through the model. In this way, the system can contribute to an explicit design rationale \cite{10.1145/1182475.1182511} while remaining an additional reflection step before empirical evaluation with real users.

\section{Formative Pilot Evaluation}
\label{sec:user-study}

We conducted an exploratory pilot evaluation with \(N=10\) participants with experience in software development, web development, or UX/UI design. The evaluation examined whether prospective users could work with and interpret the NeuroDivSim workflow and aimed to identify opportunities for prototype refinement. It was conducted with an earlier version of the prototype.

Participants selected a website or application, described a usage scenario in natural language, and could optionally provide an interface screenshot. They inspected the generated scenario models before running the simulation and exploring cognitive reference configurations, events, and design recommendations. Afterwards, participants completed the UEQ-S \cite{schrepp2017ueqs}, additional five-point rating-scale items on comprehensibility, usability, plausibility, and perceived utility, and open-ended questions.

Participants rated both understanding the system's basic purpose and completing the intended tasks at (M=4.80) ((SD=0.42)) on the five-point scales. Modeled differences between cognitive reference configurations received (M=4.70) ((SD=0.48)). Recommendations and potential use formed the highest-rated aggregated category ((M=4.53), (SD=0.32)); result presentation and findability received the lowest ((M=4.00), (SD=0.68)). The UEQ-S showed a positive descriptive tendency, with a pragmatic mean of (M=1.65), a hedonic mean of (M=2.20), and an overall mean of (M=1.92). Given the small formative sample, these values are descriptive and are not interpreted as a standardized benchmark evaluation. Open-ended responses highlighted the structured scenario models, configuration comparisons, and recommendations, while difficulties concerned interpreting metrics and events and relating them to interaction steps.

These findings informed subsequent refinement of NeuroDivSim. The result presentation was reorganized around individual interaction steps, events were connected more explicitly to affected steps, and additional explanations were introduced for model attributes, metrics, events, and recommendations. The version described in Section~\ref{sec:design} incorporates these refinements. As a formative pilot evaluation, these findings provide initial insights into how prospective users engaged with and interpreted the workflow, but do not establish the empirical validity of the simulation or its effects on design quality and accessibility outcomes.

\section{General Discussion}
\label{sec:discussion}

NeuroDivSim explores simulation as design reflection rather than synthetic participation: designers vary explicit cognitive assumptions while holding a modeled task, interface, and environment constant, then trace consequences to steps and recommendations.

\subsection{Simulation as Design Reflection}
\label{sec:reflection-layer}

The system addresses iterations between empirical evaluations, asking a restricted question: under the current assumptions, which aspects of an interaction warrant closer examination? It complements accessibility guidance and research with neurodivergent people \cite{wcag22,holmes2018mismatch,10.1145/3613904.3642424}, contextualizing possible responses by linking them to task steps and triggering assumptions. Controlled comparisons realize the counterfactual potential of computational models \cite{10.1145/3491102.3517739}.

Unlike SimUser, where a generative agent acts and produces feedback \cite{10.1145/3613904.3642481}, NeuroDivSim uses an LLM to construct a revisable scenario representation. Deterministic downstream states, events, and recommendations keep comparisons reproducible and traceable. This trace supports reflective design and design rationale \cite{sengers2005reflective,10.1145/1182475.1182511}: the system exposes why an interaction became a candidate for reconsideration. Consistent with human--AI guidance, designers retain the ability to inspect and correct the system's interpretation \cite{amershi2019guidelines}.

\subsection{Inspectability and Boundaries of Cognitive Modeling}
\label{sec:modeling-variation}

Any computational representation of cognitive diversity abstracts. The generic, ADHD-informed, and dyslexia-informed configurations neither represent individuals nor homogeneous populations. Consistent with neurodiversity and ability-based perspectives \cite{chapman2021neurodiversity,10.1145/3711106,wobbrock2011ability}, they operate on shared dimensions such as reading difficulty, attention stability, working-memory stability, and distraction sensitivity. Diagnosis-informed presets are provisional entry points, not behavioral specifications; future configurations should increasingly be organized around interaction-relevant characteristics.

Current values, weights, thresholds, and relationships are heuristic and uncalibrated. Their precision denotes reproducible output, not measured probability or empirical validity. Separating components makes them inspectable and replaceable, but transparency alone does not validate them. Problem formulation and abstraction embed consequential choices before computation begins \cite{passi2019problem,selbst2019fairness}; the system therefore cannot model lived experience and must remain alongside participatory and empirical methods.

The same boundary addresses known harms of disability simulation \cite{bennett2019empathy,silverman2015stumbling}. NeuroDivSim should not claim empathy, certify accessibility, rank groups, or replace research with neurodivergent people. Diagnosis-labeled presets, precise-looking scores, and recommendations may amplify stereotypes, overreliance, or ``accessibility theater''; explanations alone do not prevent automation bias \cite{bucinca2021trust}. Screenshots may also contain sensitive information. Labels, documentation, and studies should communicate these constraints and test users' calibration.

\subsection{Limitations and Future Work}
\label{sec:limitations}

The pilot shows that prospective users could engage with the workflow and motivated clearer links among steps, events, attributes, metrics, and recommendations. It does not establish effects on decisions or interface quality, nor correspondence with observed behavior.

Evaluation should therefore separate two questions. \textit{Design utility} requires comparison with practices such as guidelines or cognitive walkthroughs, examining concerns, alternatives, and rationale. \textit{Model validity} requires testing relationships, sensitivity to parameters, and calibration with neurodivergent participants. Prior computational-rationality work demonstrates the value of combining simulation with human studies while explicitly confronting transfer to realistic contexts \cite{lingler2026multitasking}. A reflection tool may aid reasoning without predicting behavior, while calibrated relationships alone do not establish practical utility; both model and use require evaluation.

\section{Conclusion}
\label{sec:conclusion}

This work asked whether simulation can support designers in considering cognitive diversity without requiring a generative agent to act as a surrogate for a human user. \textit{NeuroDivSim} provides one approach to this question by shifting the role of simulation from representing a user toward examining explicit assumptions about an interaction. It combines LLM supported formalization of a usage situation with inspectable cognitive reference configurations and deterministic simulation, allowing modeled consequences to be traced without presenting them as accounts of human experience.

The resulting workflow positions \textit{NeuroDivSim} as an additional reflection step between existing knowledge about users and subsequent empirical evaluation. Its formative pilot informed the refinement of this workflow, while questions of design utility and model validity remain for future investigation. \textit{NeuroDivSim} therefore illustrates a complementary role for user simulation in HCI: not to replace engagement with neurodivergent people, but to make assumptions about cognitive diversity explicit and available for examination while design decisions are still being made.

\clearpage

\bibliographystyle{ACM-Reference-Format}
\bibliography{base}

\clearpage

\appendix

\section{NeuroDivSim Interaction Workflow}
\label{app:interaction-workflow}

\begin{figure}[H]
    \centering
    \includegraphics[
        width=\textwidth,
        height=0.82\textheight,
        keepaspectratio
    ]{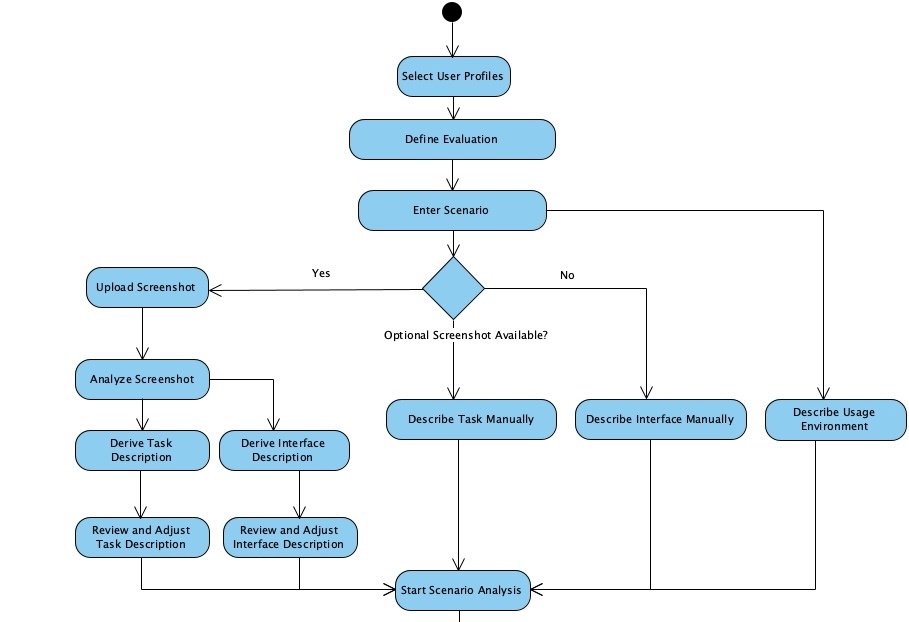}
    \caption{NeuroDivSim interaction workflow (Part 1 of 2).}
    \Description{The first part of the interaction workflow shows the initial
    stages of NeuroDivSim, including scenario specification, optional
    screenshot analysis, and the construction and review of task and
    interface descriptions.}
    \label{fig:interaction-workflow-1}
\end{figure}

\begin{figure}[H]
    \centering
    \includegraphics[
        width=\textwidth,
        height=0.82\textheight,
        keepaspectratio
    ]{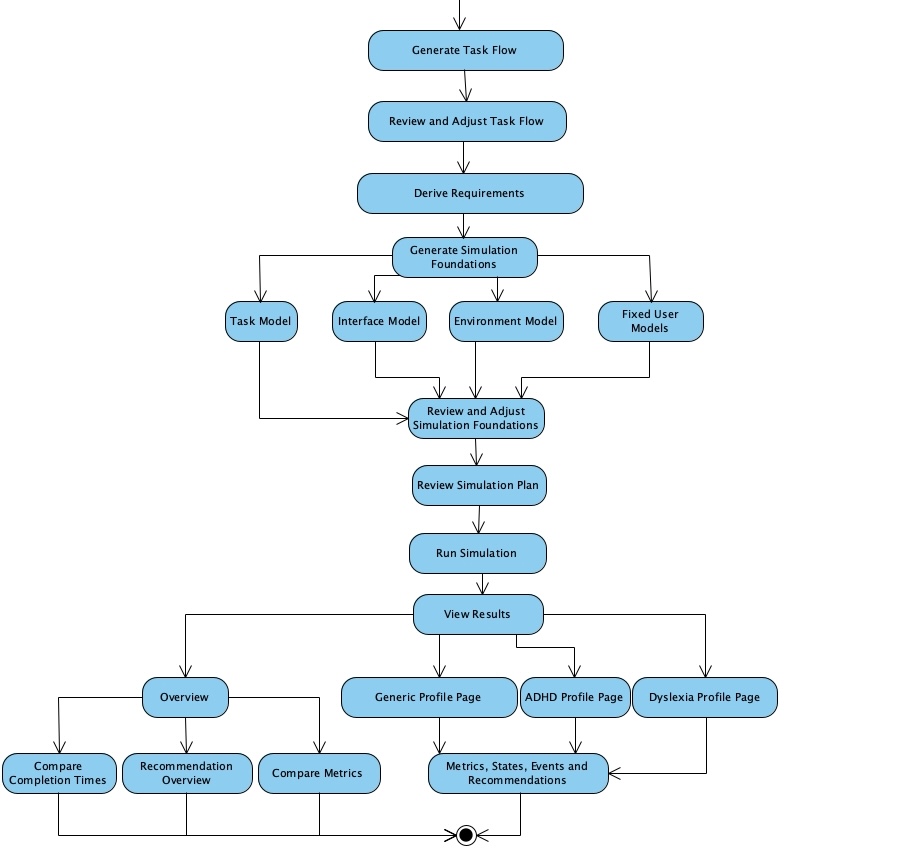}
    \caption{NeuroDivSim interaction workflow (Part 2 of 2).}
    \Description{The second part of the interaction workflow shows task flow
    generation, model construction and review, simulation, and the subsequent
    exploration of metrics, states, events, comparisons, and design
    recommendations.}
    \label{fig:interaction-workflow-2}
\end{figure}

Figures~\ref{fig:interaction-workflow-1} and
\ref{fig:interaction-workflow-2} provide a detailed view of the complete
interaction workflow in NeuroDivSim, from scenario specification and model
construction to simulation and result exploration.

\clearpage

\clearpage
\section{NeuroDivSim Interface Examples}
\label{app:interface-examples}

The following screenshots illustrate selected stages of the NeuroDivSim
workflow using a hotel-booking scenario. They are intended to provide an
impression of the implemented interface rather than a complete step-by-step
walkthrough.

\begin{figure}[H]
    \centering
    \includegraphics[
        width=\textwidth,
        height=0.78\textheight,
        keepaspectratio
    ]{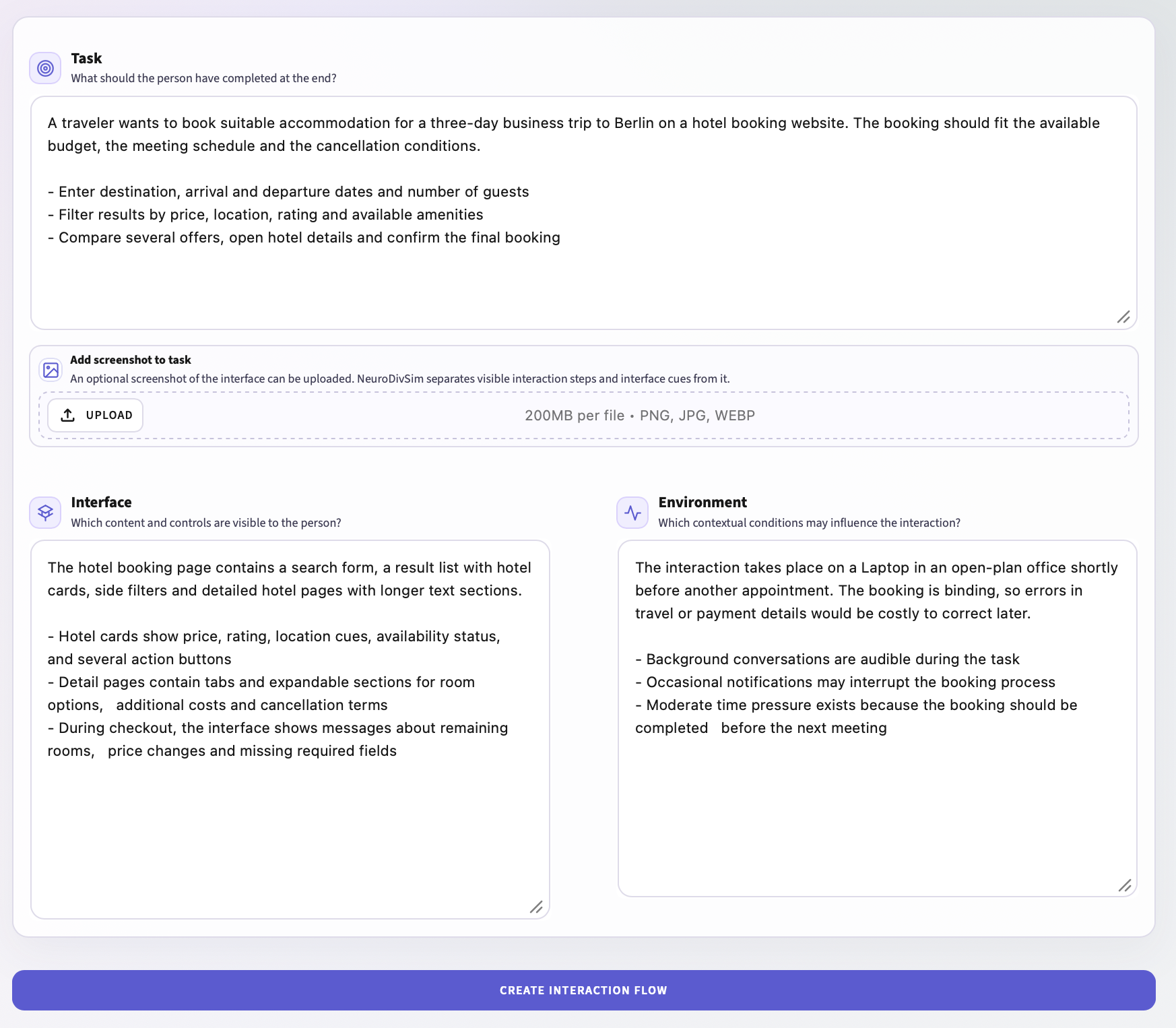}
    \caption{Example scenario specification in NeuroDivSim.}
    \Description{The NeuroDivSim interface showing the specification of a
    hotel-booking task together with descriptions of the interface and
    environmental context.}
    \label{fig:example-scenario}
\end{figure}

\clearpage

\begin{figure}[H]
    \centering
    \includegraphics[
        width=\textwidth,
        height=0.78\textheight,
        keepaspectratio
    ]{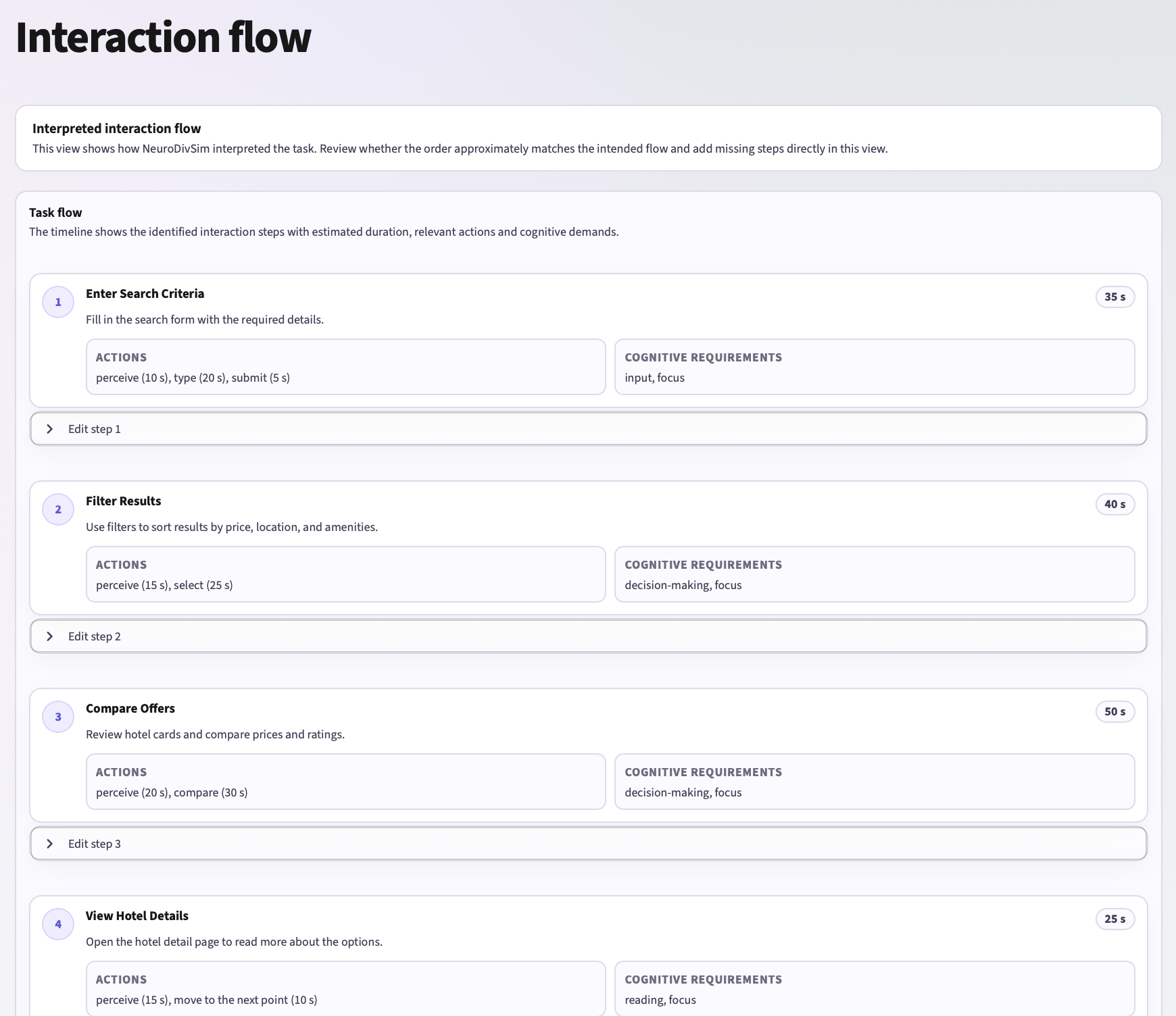}
    \caption{Example of an interpreted interaction flow.}
    \Description{The interaction flow generated for the hotel-booking scenario,
    showing individual task steps together with estimated durations, actions,
    and cognitive requirements.}
    \label{fig:example-interaction-flow}
\end{figure}

\clearpage

\begin{figure}[H]
    \centering
    \includegraphics[
        width=\textwidth,
        height=0.78\textheight,
        keepaspectratio
    ]{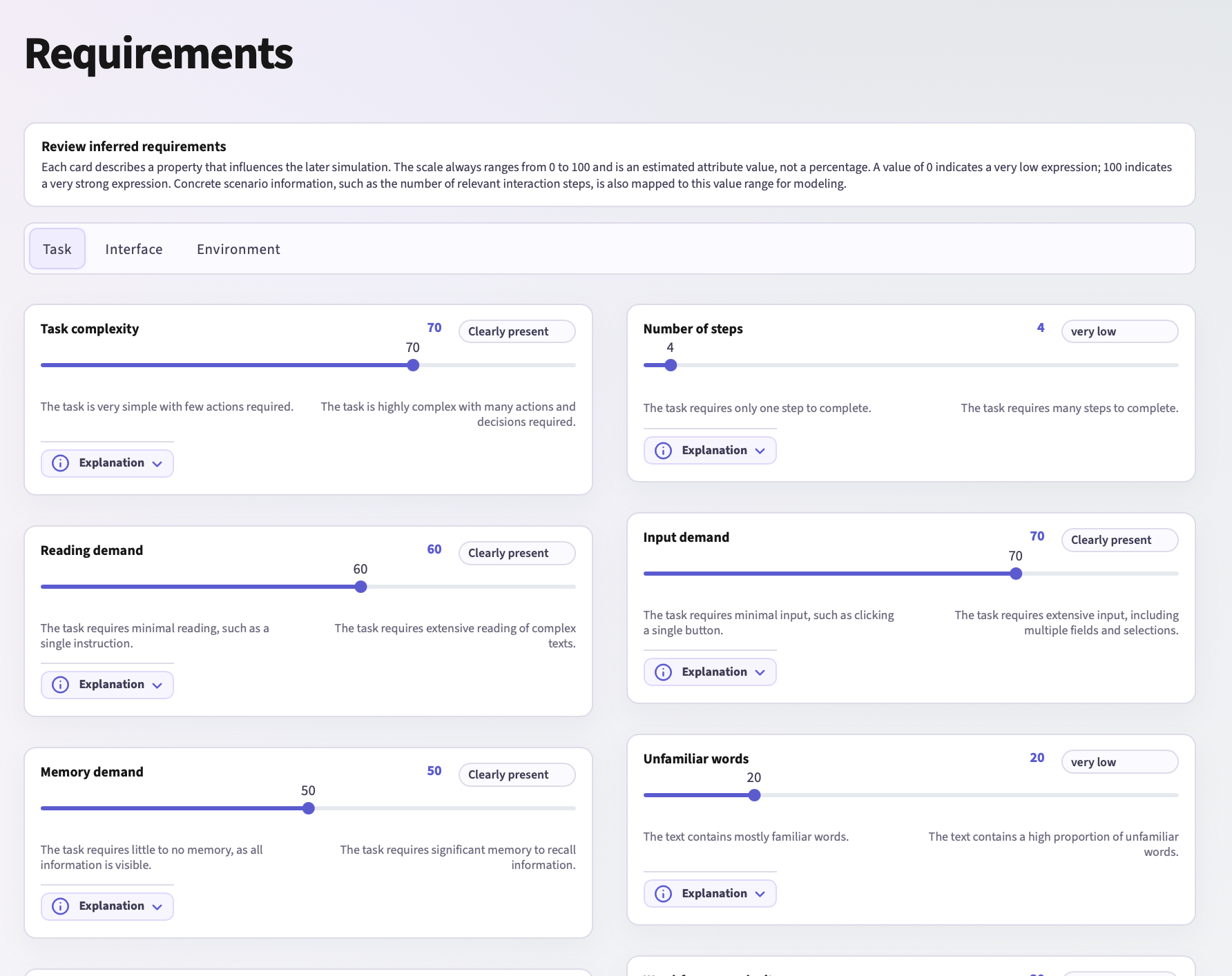}
    \caption{Example of inspectable model requirements.}
    \Description{The requirements view showing editable task attributes on a
    common scale, including task complexity, number of steps, reading demand,
    input demand, memory demand, and unfamiliar words.}
    \label{fig:example-requirements}
\end{figure}

\clearpage

\begin{figure}[H]
    \centering
    \includegraphics[
        width=\textwidth,
        height=0.78\textheight,
        keepaspectratio
    ]{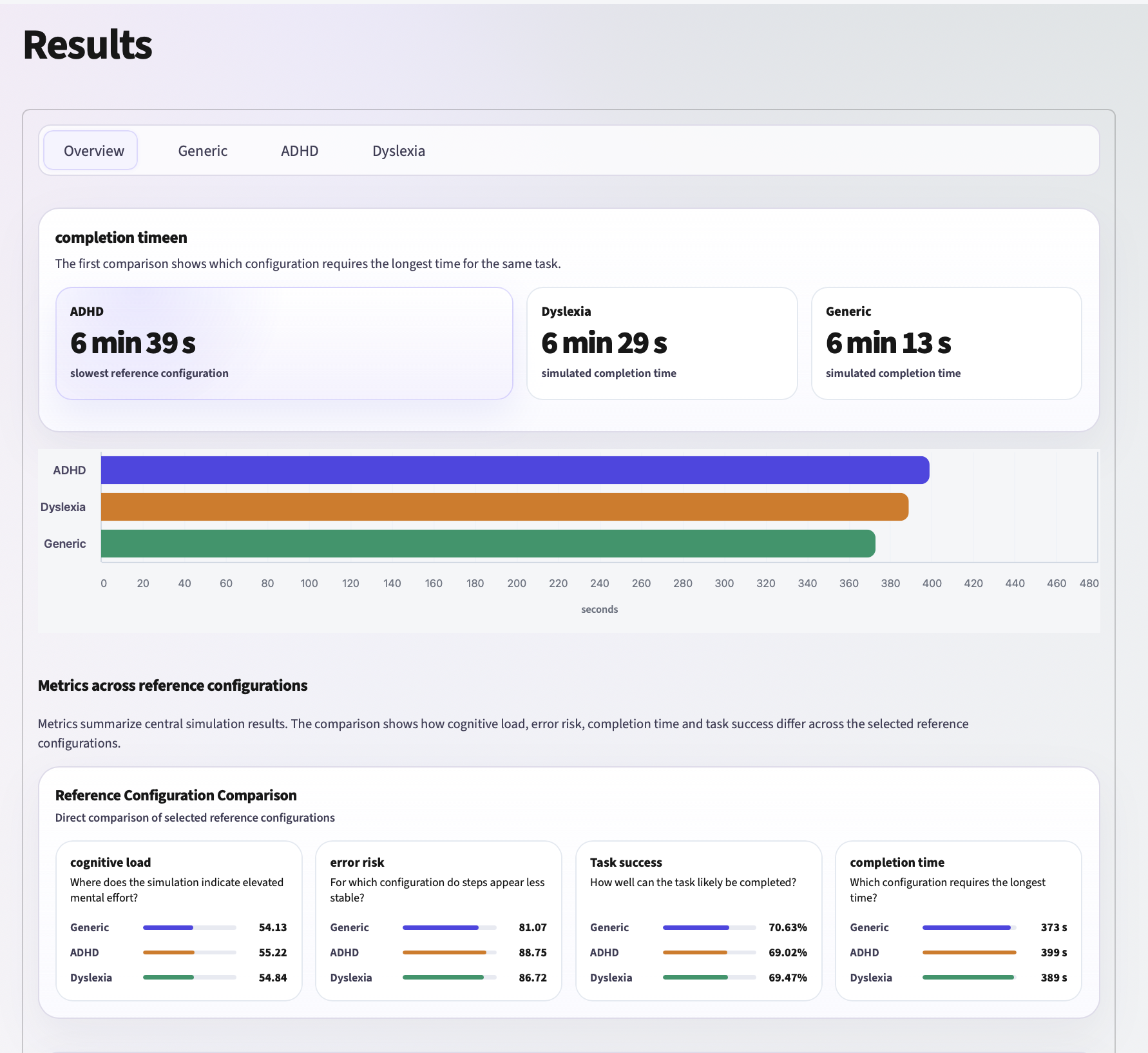}
    \caption{Example overview of simulation results.}
    \Description{The results overview comparing the Generic, ADHD, and Dyslexia
    reference configurations in terms of simulated completion time, cognitive
    load, error risk, and task success.}
    \label{fig:example-results}
\end{figure}

\clearpage

\begin{figure}[H]
    \centering
    \includegraphics[
        width=\textwidth,
        height=0.82\textheight,
        keepaspectratio
    ]{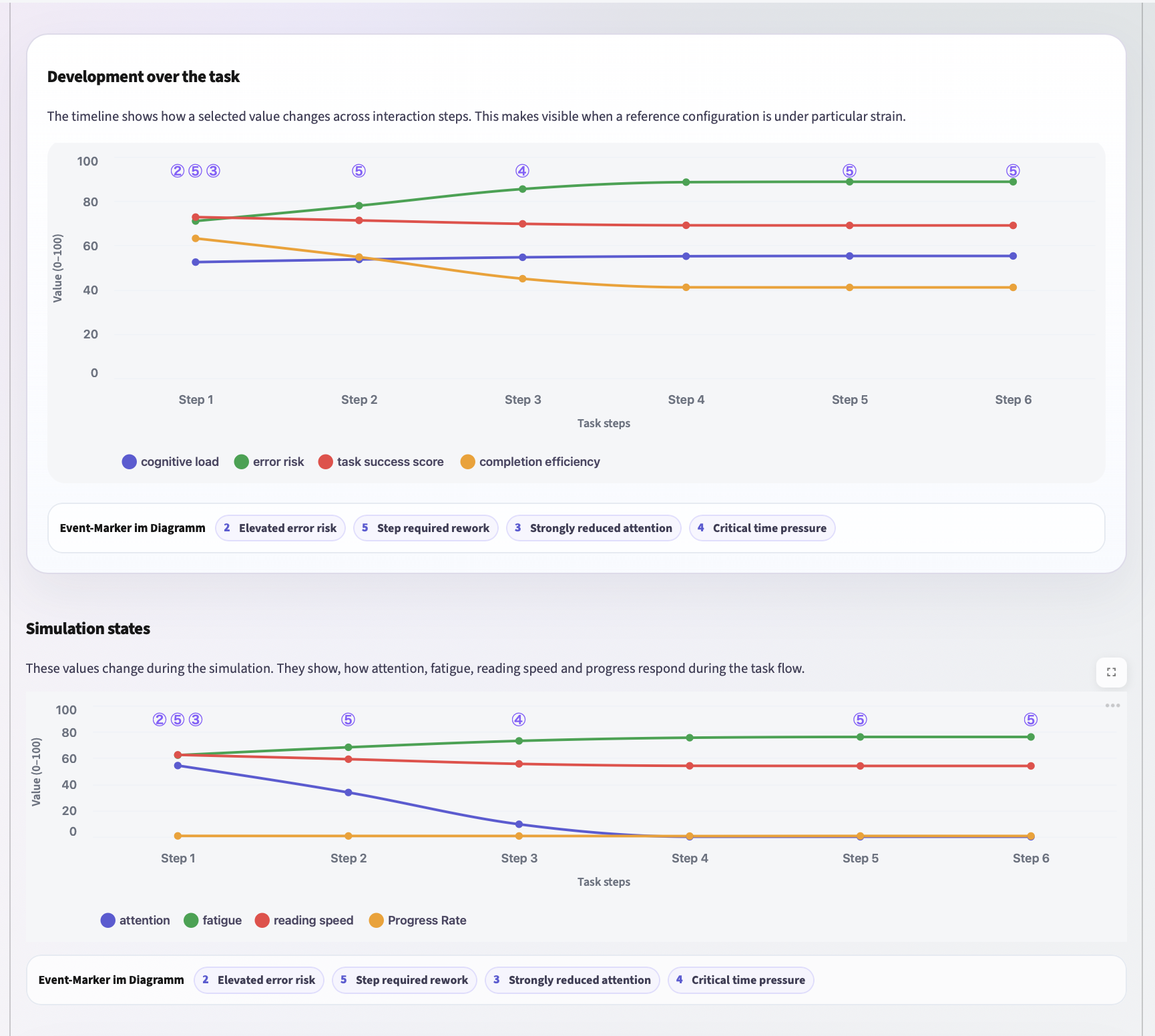}
    \caption{Detailed simulation results for an ADHD-informed reference configuration.}
    \Description{NeuroDivSim result view showing the development of simulation
    metrics and dynamic states across six interaction steps. The upper chart
    displays cognitive load, error risk, task success score, and completion
    efficiency. The lower chart displays attention, fatigue, reading speed,
    and progress rate. Event markers connect threshold-based simulation events
    to the task steps at which they occurred.}
    \label{fig:example-simulation-results}
\end{figure}

\clearpage

\begin{figure}[H]
    \centering
    \includegraphics[
        width=\textwidth,
        height=0.78\textheight,
        keepaspectratio
    ]{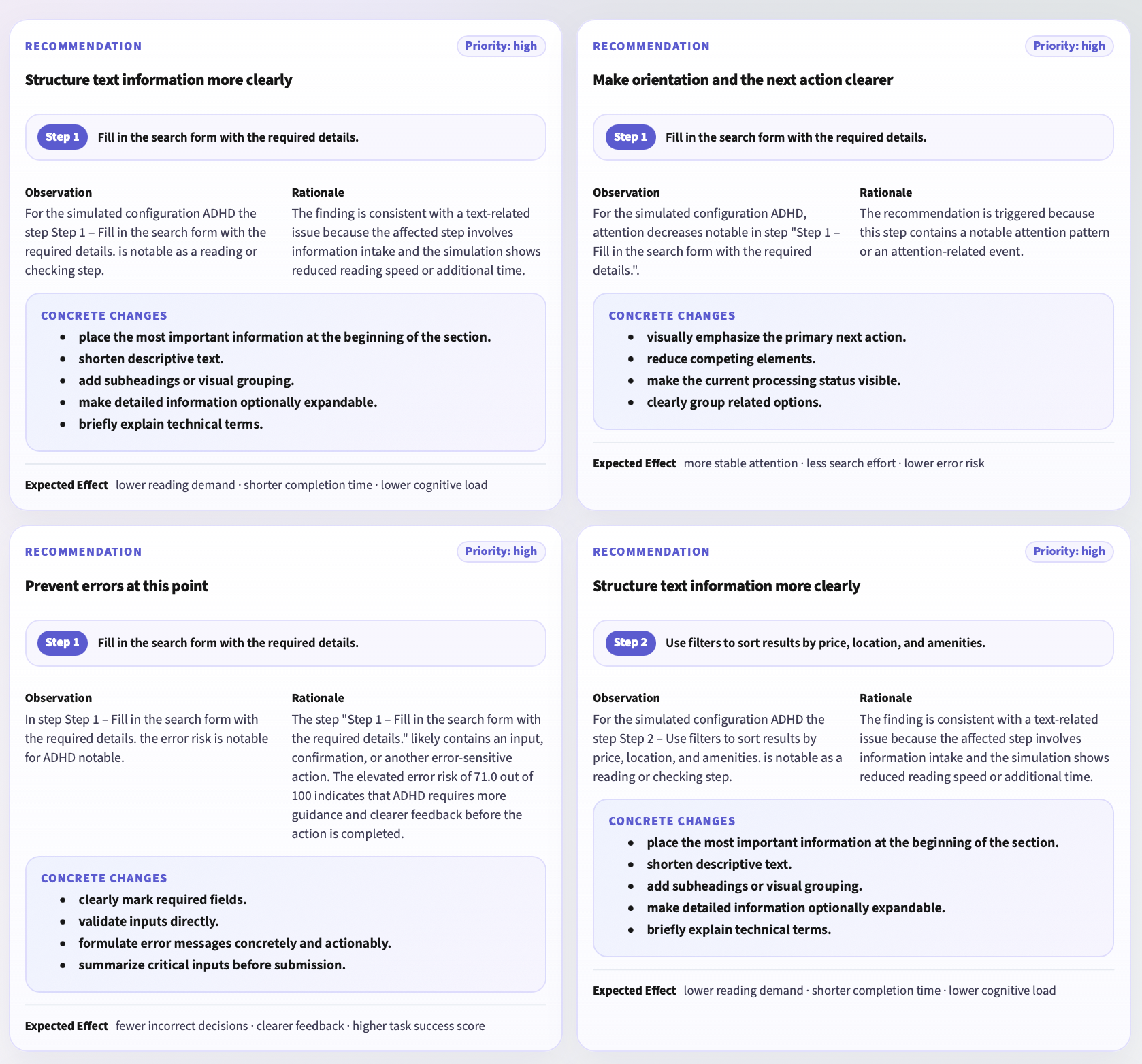}
    \caption{Example design recommendations for the ADHD-informed reference configuration.}
    \Description{The recommendation view for the ADHD-informed reference
    configuration, showing high-priority recommendations associated with
    individual interaction steps, together with the corresponding observations,
    rationales, concrete changes, and expected effects.}
    \label{fig:example-recommendations}
\end{figure}

\end{document}